\documentclass[pra,aps,twocolumn,nobalancelastpage,superscriptaddress,colorinlistoftodos]{revtex4-2}

\usepackage[utf8]{inputenc}
\usepackage[T1]{fontenc}
\usepackage[english]{babel}
\usepackage{csquotes}
\usepackage{comment}

\usepackage{amsmath}
\usepackage{amsfonts}
\usepackage{relsize}
\usepackage{bm}
\usepackage{braket}

\usepackage[margin=2cm]{geometry}

\usepackage{siunitx}
\DeclareSIUnit{\bits}{bits}
\DeclareSIUnit{\dBc}{dBc}
\DeclareSIUnit{\dBm}{dBm}
\DeclareSIUnit{\sample}{Sa}
\DeclareSIUnit{\snu}{SNU}
\DeclareSIUnit{\baud}{Baud}

\usepackage{graphicx}
\graphicspath{{./figures/}}
\usepackage{pgf}
\usepackage{float}

\everymath=\expandafter{\the\everymath\displaystyle}
\usepackage[caption=false]{subfig}

\usepackage{array}
\usepackage{multirow}
\usepackage{makecell}
\usepackage{longtable}
\usepackage{booktabs}

\usepackage[hidelinks]{hyperref}
\usepackage{natbib}
\usepackage[normalem]{ulem} 
\usepackage{glossaries-extra}
\setabbreviationstyle[acronym]{long-short}

\newabbreviation{qkd}{QKD}{quantum key distribution}
\newabbreviation{dvqkd}{DV-QKD}{discrete-variable quantum key distribution}
\newabbreviation{cvqkd}{CV-QKD}{continuous-variable quantum key distribution}
\newabbreviation{dsp}{DSP}{digital signal processing}
\newabbreviation{dac}{DAC}{digital-to-analog converter}
\newabbreviation{adc}{ADC}{analog-to-digital converter}
\newabbreviation{pbs}{PBS}{polarizing beam splitter}
\newabbreviation{lo}{LO}{local oscillator}
\newabbreviation{qci}{QCI}{quantum communication infrastructure}
\newabbreviation{mbc}{MBC}{modulator bias controller}
\newabbreviation{voa}{VOA}{variable optical attenuator}
\newabbreviation{skr}{SKR}{secret key rate}
\newabbreviation{cw}{CW}{continuous-wave}
\newabbreviation{fpga}{FPGA}{field-programmable gate array}
\newabbreviation{mpc}{MPC}{motorized polarization controller}
\newabbreviation{dwdm}{DWDM}{dense wavelength-division multiplexing}
\newabbreviation{wdm}{WDM}{wavelength-division multiplexing}
\newabbreviation{tdc}{TDC}{time-to-digital converter}
\newabbreviation{snr}{SNR}{signal-to-noise ratio}
\newabbreviation{qber}{QBER}{quantum bit error rate}
\newabbreviation{bpd}{BPD}{balanced photodetector}
\newabbreviation{smf}{SMF}{single mode fiber}
\newabbreviation{mls}{MLS}{maximum-length sequence}

\usepackage{todonotes}
\definecolor{edo}{RGB}{0, 150, 0}

\definecolor{yoann}{RGB}{52, 152, 219}

\definecolor{mattia}{HTML}{FF8000}

\definecolor{pino}{RGB}{52, 219, 52}

\usepackage[dvipsnames]{xcolor}

\date{}

\begin{document}
\title{Multiplexing of Continuous-Variable and Discrete-Variable Quantum Key Distribution Systems over Fibered and Free-Space Channels}

\author{Mattia Sabatini}
\affiliation{Università degli Studi di Padova, Dipartimento di Ingegneria dell’Informazione, via Gradenigo 6B, IT-35131 Padova, Italy}
\author{Edoardo Rossi}
\affiliation{Università degli Studi di Padova, Dipartimento di Ingegneria dell’Informazione, via Gradenigo 6B, IT-35131 Padova, Italy}
\author{Matías R. Bolaños}
\affiliation{Università degli Studi di Padova, Dipartimento di Ingegneria dell’Informazione, via Gradenigo 6B, IT-35131 Padova, Italy}
\author{Francesco Vedovato}
\affiliation{Università degli Studi di Padova, Dipartimento di Ingegneria dell’Informazione, via Gradenigo 6B, IT-35131 Padova, Italy}
\affiliation{Padua Quantum Technologies Research Center, Università degli Studi di Padova, via Gradenigo 6B, IT-35131 Padova, Italy}
\author{Thomas Liege}
\affiliation{Sorbonne Université, CNRS, LIP6, F-75005 Paris, France}
\affiliation{ONERA, DOTA, Paris Saclay University, F-92322 Châtillon, France}
\author{Eleni Diamanti}
\affiliation{Sorbonne Université, CNRS, LIP6, F-75005 Paris, France}
\author{Giuseppe Vallone}
\affiliation{Università degli Studi di Padova, Dipartimento di Ingegneria dell’Informazione, via Gradenigo 6B, IT-35131 Padova, Italy}
\affiliation{Padua Quantum Technologies Research Center, Università degli Studi di Padova, via Gradenigo 6B, IT-35131 Padova, Italy}
\author{Paolo Villoresi}
\affiliation{Università degli Studi di Padova, Dipartimento di Ingegneria dell’Informazione, via Gradenigo 6B, IT-35131 Padova, Italy}
\affiliation{Padua Quantum Technologies Research Center, Università degli Studi di Padova, via Gradenigo 6B, IT-35131 Padova, Italy}
\author{Yoann Piétri}
\affiliation{Università degli Studi di Padova, Dipartimento di Ingegneria dell’Informazione, via Gradenigo 6B, IT-35131 Padova, Italy}
\affiliation{Sorbonne Université, CNRS, LIP6, F-75005 Paris, France}
\author{Marco Avesani}
\email{marco.avesani@unipd.it}
\affiliation{Università degli Studi di Padova, Dipartimento di Ingegneria dell’Informazione, via Gradenigo 6B, IT-35131 Padova, Italy}
\affiliation{Padua Quantum Technologies Research Center, Università degli Studi di Padova, via Gradenigo 6B, IT-35131 Padova, Italy}
\date{\today}

\begin{abstract}
Future quantum communication infrastructures will need to serve 
heterogeneous users on shared physical channels: short-range, 
high-throughput links favor Continuous-Variable Quantum Key Distribution 
(CV-QKD), while long-reach, high-loss links remain the domain of 
Discrete-Variable QKD (DV-QKD). Wavelength-division multiplexing (WDM) 
of the two protocols on a common channel would address both regimes 
simultaneously, but their markedly different noise sensitivities make 
coexistence non-trivial and, to date, experimentally untested. Here we 
report the first simultaneous operation of two independent CV- and 
DV-QKD systems on a common optical channel, using standard  C-band DWDM filters at \SI{1550.12}{\nano\meter} 
(CV) and \SI{1545.32}{\nano\meter} (DV). We demonstrate joint operation 
on both optical fiber and a \SI{620}{\meter} urban daylight free-space 
link. On fiber, the two systems exhibit the 
expected complementarity, crossing over at \SI{7.56}{\decibel} of 
channel loss where both deliver ${\sim}1.43\,\mathrm{Mbit/s}$; in 
daylight free-space, both sustain Mbit/s key rates under time-varying 
atmospheric attenuation. Across all configurations we observe no 
measurable multiplexing-induced penalty in QBER or excess noise. These 
results establish hybrid CV--DV WDM as a practical building block for 
heterogeneous quantum communication networks, where metropolitan high-throughput users and long-reach
backbone links can be served on a single physical infrastructure.
\end{abstract}

\maketitle

\section{Introduction}

\Gls{qkd} enables the exchange of cryptographic keys with a security guaranteed by the fundamental principles of quantum mechanics, and therefore independent of any computational assumptions~\cite{scarani_security_2009, gisin_quantum_2002}.
Over the past two decades, it has become a mature field of research and a wide range of QKD implementations has been demonstrated, exploiting various degrees of freedom of the electromagnetic field to encode quantum information, which are regrouped in two families~\cite{xu_secure_2020, pirandola_advances_2020}.
In discrete-variable (DV) schemes, information is carried by properties assuming discrete values with measurements yielding discrete outcomes, typically polarization, time-bin, phase, or orbital-angular-momentum~\cite{bennett_quantum_2014, ekert_quantum_1991}.
In contrast, continuous-variable (CV) protocols encode information in the field quadratures of coherent, squeezed or thermal states and recover it using coherent detection techniques yielding continuous results~\cite{grosshans_continuous_2002, weedbrook_gaussian_2012, laudenbach_continuous-variable_2018, zhang_continuous-variable_2024}.

Operating using different encodings, DV and \gls{cvqkd} have both advantages and inconvenient: \gls{cvqkd} can deliver high \glspl{skr} at low-to-moderate loss, whereas \gls{dvqkd} is typically more robust in the high-loss regime, where CV performance is limited by low \gls{snr}.
Indeed, \gls{dvqkd} systems have proven their effectiveness in long-distance key distribution, supported by well-established security proofs, and have become the benchmark for high-loss channels and network backbones~\cite{boaron_secure_2018, liu_experimental_2023}. 
On the other hand, \gls{cvqkd} systems excel in metropolitan-scale networks, achieving higher key rates at short distances and offering lower cost and greater potential for integration~\cite{hajomer_continuous-variable_2024, pietri_experimental_2024, aldama_integrated_2025, peri_high-performance_2025}.
A natural way to exploit this complementarity is to let CV and DV systems share the same optical path via \gls{wdm}, a standard approach in classical optical networks. However, coexistence is not guaranteed \textit{a priori}: the two systems have different noise sensitivities, so multiplexing may introduce cross-talk and additional noise, which translates into \gls{qber} penalty for \gls{dvqkd} or excess-noise penalty for \gls{cvqkd}. In this work, we experimentally investigate this coexistence by operating two fully independent \gls{cvqkd} and \gls{dvqkd} systems at distinct C-band wavelengths simultaneously on the same optical link.


Recent field deployments show that \gls{qkd} is evolving from isolated point-to-point links into large-scale networks. 
Early metropolitan testbeds such as the SECOQC network in Vienna and the Tokyo \gls{qkd} Network already demonstrated multi-vendor, multi-protocol operation over reconfigurable topologies~\cite{peev_secoqc_2009, sasaki_field_2011}. 
The China Quantum Communication Network (CN-QCN) now spans more than \SI{10000}{\kilo\meter} of fiber and dozens of backbone and metropolitan nodes, operating as an integrated trusted-relay infrastructure~\cite{chen_carrier_grade_2025}. 
At the metropolitan scale, the heterogeneous SDN-controlled MadQCI (Madrid Quantum Communication Infrastructure) further demonstrates that multi-vendor \gls{qkd} systems can be integrated on shared telecom infrastructure~\cite{martin_madqci_2024}.

From a network perspective, a hybrid CV--DV architecture naturally fits the heterogeneous requirements of a realistic \gls{qci}. 
In such a scenario, short-range, high-throughput users (for instance metropolitan data centers or aggregation nodes) can be served by \gls{cvqkd}, which offers higher secret key rates at moderate loss. 
In parallel, long-reach or high-loss users can be connected via \gls{dvqkd} on the same physical infrastructure, benefiting from its superior loss tolerance. A wavelength-division-multiplexed CV--DV link therefore allows a network operator to address both use cases simultaneously, without duplicating fibers or free-space terminals.

Our experiment has been designed to reflect this kind of asymmetric deployment. The CV transmitter and receiver are positioned closer to each other, while the DV transmitter and receiver are housed in a separate laboratory and connected via additional fiber links.

To the best of our knowledge, this is the first experimental demonstration of wavelength-division multiplexing between independent \gls{cvqkd} and \gls{dvqkd} systems operating simultaneously on the same optical link. We validate coexistence over a 20-km fiber spool, through a channel implemented with a variable optical attenuator (VOA), and a $\sim$600 m free-space link, demonstrating simultaneous operation of both systems under realistic conditions. For the free space link, the usage of the local local oscillator (LLO) technique for the \gls{cvqkd} is also a noticeable feature as such systems have only been demonstrated on emulated free space links~\cite{liao_high-rate_2025}.

The rest of the paper is organized as follows: in section~\ref{sec:hardware} the CV and \gls{dvqkd} hardware is described, along with an analysis of the multiplexing on their performance, in section~\ref{sec:fiber-results}, the results of the multiplexing over fiber channels are presented, followed by the results over the free space channel in section~\ref{sec:free-space-results}. Finally, the work is concluded in section~\ref{sec:conclusion}.

\section{Hardware description\label{sec:hardware}}

\subsection{CV-QKD system}

\begin{figure*}
    \centering
    \includegraphics[scale=0.055]{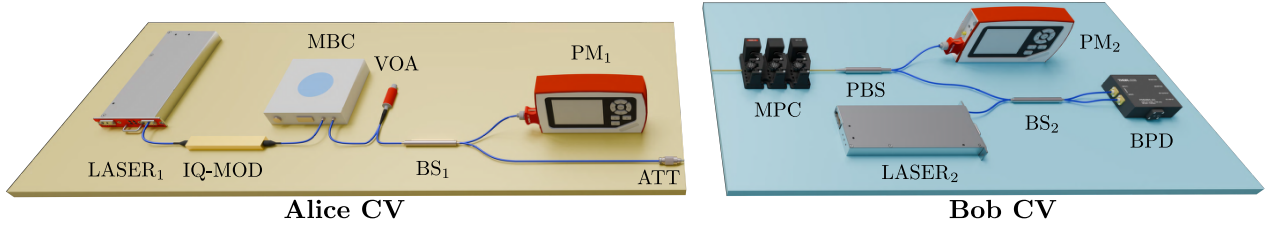}
    \caption{Experimental setup of the CV-QKD system. At Alice, a narrow-linewidth \gls{cw} laser ($\mathrm{LASER}_1$) is modulated by an IQ modulator ($\mathrm{IQ}\text{-}\mathrm{MOD}$), attenuated to the quantum level by a variable optical attenuator (VOA) and a fixed attenuator (ATT). A 95:5 beam splitter ($\mathrm{BS}_1$) sends 95\% of the power to an optical power meter ($\mathrm{PM}_1$) for monitoring, while the remaining 5\% constitutes the quantum signal sent towards the channel. At Bob, the signal passes through a motorized polarization controller (MPC), a \gls{pbs} and an optical power meter ($\mathrm{PM}_2$), which are used together to align the polarization of the receiver with that of the transmitter, is then interfered with an independent \gls{lo} ($\mathrm{LASER}_2$) on a 50:50 beam splitter ($\mathrm{BS}_2$), and the outputs are measured with a balanced photodetector (BPD) for coherent detection and subsequent digital processing using the QOSST software~\cite{Pietri2024}.}
    \label{fig:cv-qkd_setup}
\end{figure*}

The \gls{cvqkd} system used in this work is illustrated in Fig.~\ref{fig:cv-qkd_setup} and implements a Gaussian-modulated coherent-state protocol, based on IQ modulation with digital filtering and coherent detection using the RF heterodyne technique.
All digital operations are done using the QOSST open source software~\cite{Pietri2024}.

Alice is composed of $\mathrm{LASER}_1$, a \gls{cw} DFB fiber laser (NKT Koheras Basik X15 centered at $\lambda_{\rm CV}=\SI{1550.12}{\nano\meter}$ (C34 of the ITU grid), emitting \SI{10}{\milli\watt} of optical power. The light is then modulated using an Exail MIXQER-LN-30 single-polarization IQ modulator ($\mathrm{IQ}\text{-}\mathrm{MOD}$). Part of the light is detected by an Exail \gls{mbc} to lock the modulator to the IQ modulator working point. This is followed by a \gls{voa}, and a 95:5 beam splitter ($\mathrm{BS}_1$). The \SI{95}{\percent} path is used to measure the average number of photons per symbol $\langle n \rangle$ through an optical power measurement ($\mathrm{PM}_1$), and the \SI{5}{\percent} path undergoes a final \SI{10}{\dB} optical attenuator (ATT) before reaching the output of Alice. All the fibers are polarization-maintaining. The IQ modulator is driven by a Teledyne SDR14 Tx \gls{dac}, which has a sampling rate of $\SI{2}{\giga\sample\per\second}$ and peak-to-peak amplitude of \SI{1}{\volt}. The signal waveform is generated at a symbol rate $R_s=\SI{100}{\mega\baud}$.

At Bob's side, the signal first undergoes a polarization recovery stage, composed of a Thorlabs MPC320 \gls{mpc}, a \gls{pbs} and an optical power meter ($\mathrm{PM}_2$). Then the signal is mixed in a 50:50 beam splitter ($\mathrm{BS}_2$) with the \gls{lo}, provided by another NKT Koheras Basik X15 \gls{cw} laser ($\mathrm{LASER}_2$) centered at \SI{1550.12}{\nano\meter}. After the mixing, the two paths are detected using a Thorlabs PDB480C-AC \gls{bpd} , and the resulting electrical signal is digitized by a Teledyne ADQ32 \gls{adc} operated at \SI{2.5}{\giga\sample\per\second}.

The digitized heterodyne data is stored and processed offline by the receiver \gls{dsp} chain described in Appendix~\ref{app:dsp}. No external reference clock is distributed between Alice and Bob, the sampling-clock mismatch is recovered from the two pilot tones and digitally compensated.

\subsection{DV-QKD system}

\begin{figure*}
    \centering
    \includegraphics[scale=0.055]{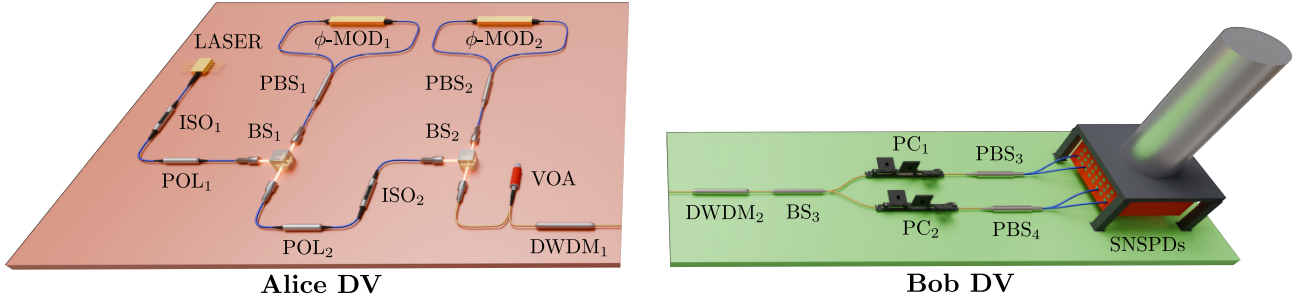}
    \caption{Experimental setup of the DV-QKD system. The transmitter includes a gain-switched DFB laser followed by an optical isolator ($\mathrm{ISO}_1$) and a polarizer ($\mathrm{POL}_1$) preparing the field in the diagonal state $\ket{D}$. Two lithium-niobate phase modulators ($\phi\text{-}\mathrm{MOD}_1$ and $\phi\text{-}\mathrm{MOD}_2$) placed within a Sagnac loop, each with identical architecture implement intensity and polarization modulation, respectively. In the first Sagnac, a $\ket{D}$-aligned polarizer ($\mathrm{POL}_2$) at the output converts the phase shift into intensity modulation for decoy-state generation, while the second Sagnac performs the polarization encoding. The modulated pulses are coupled into a standard \gls{smf} and directed into the quantum channel. At the receiver, the incoming signal is passively split by a fiber beam splitter ($\mathrm{BS}_3$) and analyzed using polarization controllers ($\mathrm{PC}_1$ and $\mathrm{PC}_2$), polarizing beam splitters ($\mathrm{PBS}_3$ and $\mathrm{PBS}_4$), and four superconducting nanowire single-photon detectors (SNSPDs) for projections onto $\ket{L}$, $\ket{R}$, $\ket{D}$, and $\ket{A}$. Yellow fibers indicate single-mode (SM) fiber, while blue fibers indicate polarization-maintaining (PM) fiber.}
    \label{fig:dv-qkd_setup}
\end{figure*}

The experimental setup for the \gls{dvqkd} system, including both transmitter and receiver, is shown in Fig.~\ref{fig:dv-qkd_setup}.
The source implements the system previously demonstrated in Ref.~\cite{bolanos_ghz-rate_2026}, based on a gain-switched laser followed by two iPOGNAC modulation stages providing intensity and polarization modulation respectively, and realizing the three-state one-decoy BB84 protocol~\cite{grunenfelder_simple_2018}. In the present work, the same configuration is used at a different laser wavelength.

The transmitter employs a gain-switched DFB laser diode operating at $\lambda_{\rm DV}=\SI{1545.32}{\nano\meter}$ (C40 of the ITU grid), producing optical pulses with $\rm{FWHM}=\SI{61.5\pm0.5}{\pico\second}$ at a repetition rate of $R_{\mathrm{DV}}=\SI{1}{\giga\hertz}$.
The laser output first passes through an optical isolator ($\mathrm{ISO}_1$) and a linear polarizer ($\mathrm{POL}_1$), which prepare the optical field in the diagonal polarization state $\ket{D}$.
The pulses are then directed to a free-space beam splitter ($\mathrm{BS}_1$), which routes the light into a Sagnac interferometric loop. 
Within this loop, a \gls{pbs} ($\mathrm{PBS}_1$) separates the two orthogonal polarization components, while an electro-optic lithium-niobate phase modulator $\phi\text{-}\mathrm{MOD}_1$ (iXblue MPZ-LN-10), placed symmetrically within the loop, applies the desired phase difference $\Delta\phi_1$ between the two components.
After a complete round trip, the optical pulses recombine and exit $\mathrm{PBS}_1$, effectively mapping the input diagonal state into an arbitrary state belonging to the X-Y equatorial plane of the Bloch sphere.
A second $\ket{D}$-aligned polarizer ($\mathrm{POL}_2$) attenuates the optical pulses by a factor of $\cos^{2}(\Delta\phi_1)$, enabling intensity modulation for decoy-state generation.

A second Sagnac modulator with identical architecture performs the polarization encoding, the only difference being the absence of a polarizer at its output. The encoded states exiting the free-space beam splitter $\mathrm{BS}_2$ are coupled into a \gls{smf}. Before entering the quantum channel, the optical signal passes through a variable optical attenuator (VOA), which sets the mean photon number to the level required by the protocol, followed by a \SI{100}{\giga\hertz}-\gls{dwdm} band-pass filter ($\mathrm{DWDM}_1$) centered at channel C40 of the ITU grid to suppress out-of-band noise and ensure spectral purity~\cite{mao_integrating_2018}.

Both the DFB laser and the two phase modulators are driven by a control system based on a \gls{fpga} (AMD/Xilinx Zynq UltraScale+ RFSoC ZCU111), which generates the synchronized electrical signals required for pulse generation and state encoding.
To allow for the symmetric configuration within the Sagnac loop, both $\phi\text{-}\mathrm{MOD}$ were driven using the \textit{balanced modulation} technique, as introduced in Ref.~\cite{berra2025generalmodelmodulationstrategies}.

At the receiver, the incoming optical signal first passes through a \SI{100}{\giga\hertz}-\gls{dwdm} band-pass filter ($\mathrm{DWDM}_2$) centered at ITU channel C40, identical to the one used at the transmitter. The filtered signal is then passively split by a 50:50 fiber beam splitter ($\mathrm{BS}_3$), implementing the passive basis choice. Each output arm is equipped with a polarization controller ($\mathrm{PC}_1$ and $\mathrm{PC}_2$) and a \gls{pbs} ($\mathrm{PBS}_3$ and $\mathrm{PBS}_4$), enabling projections onto the polarization states $\ket{L}$, $\ket{R}$, $\ket{D}$, and $\ket{A}$. The four outputs are directed to superconducting nanowire single photon detectors (ID281 SNSPDs by ID Quantique) with an average quantum efficiency of approximately \SI{80}{\percent}. 
Detection events are time-tagged by a \gls{tdc} (quTAG qutools), and the transmitter--receiver time reference is synchronized using the Qubit4Sync algorithm, without requiring an external shared clock~\cite{calderaro_fast_2020}. Finally, the timetags are stored for offline post-processing.

\subsection{Influence of multiplexing}

A key question for a \gls{wdm} architecture is whether the simultaneous operation of two independent \gls{qkd} systems on the same physical link introduces any measurable performance penalty due to cross-talk or additional noise. This point is especially relevant because the \gls{cvqkd} implementation relies on classical signals to enable robust \gls{dsp}: each frame is preceded by a \gls{mls} used for synchronization, and two frequency-multiplexed relatively strong pilot tones are added to the transmitted waveform for clock recovery, carrier-frequency estimation, and phase recovery.
Although the DV quantum channel is filtered at the receiver by a \SI{100}{\giga\hertz} \gls{dwdm}, any out-of-band photons associated with the classical CV components, for instance due to finite filter isolation or Raman scattering generated in the fiber, could, in principle, increase the background noise and thus the DV \gls{qber}.

From the fiber-based simultaneous \gls{wdm} measurements reported in this work, up to approximately \SI{36}{\decibel} of \gls{dvqkd} channel loss, we found no measurable change in the DV \gls{qber} or in the CV estimated excess noise; accordingly, the secret key rates of both systems remain unchanged within this regime.

\begin{figure}
    \centering
    \includegraphics[width=\columnwidth]{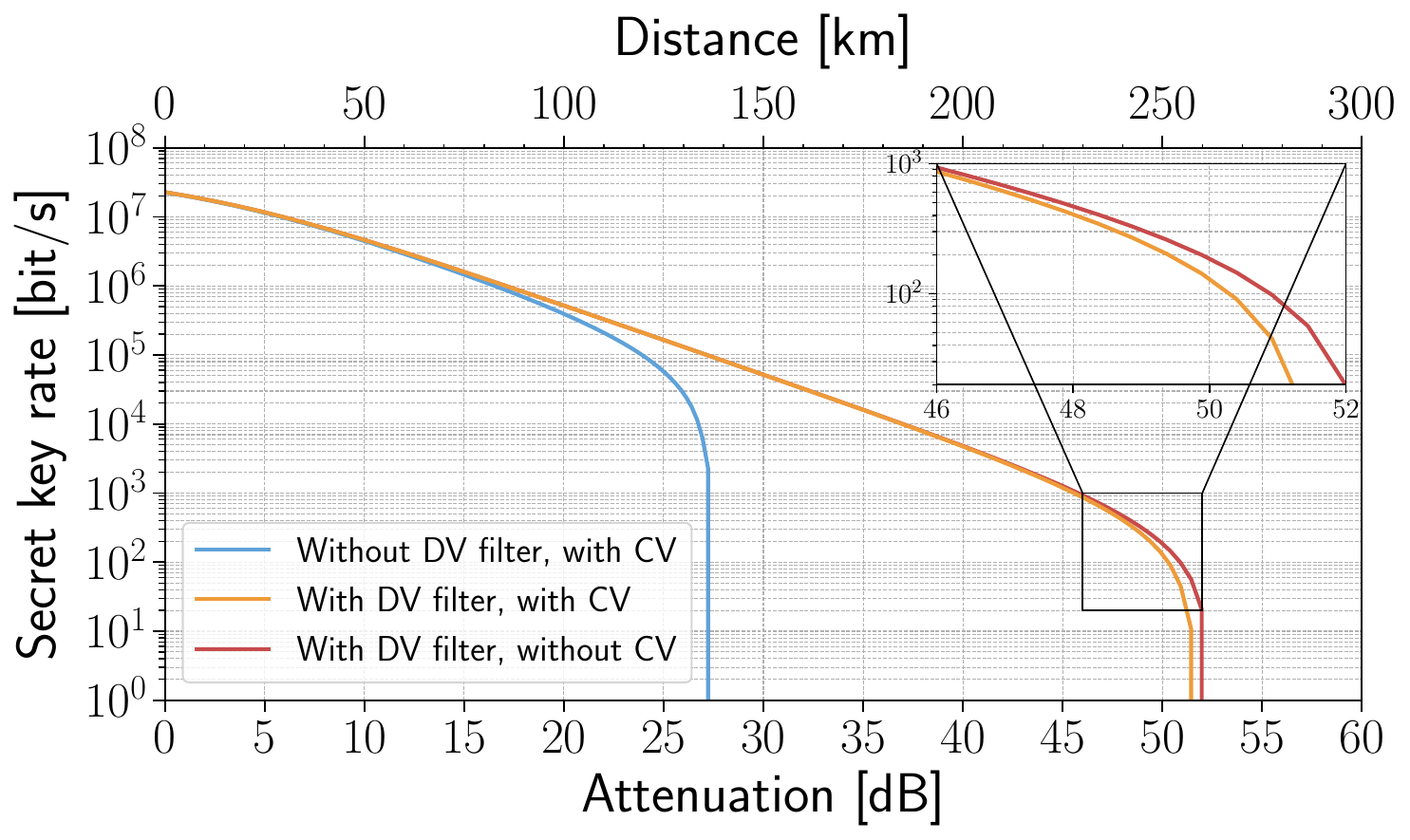}
    \caption{Simulated asymptotic DV-QKD \gls{skr} as a function of channel attenuation using background count rates measured with the DV signal blocked under different multiplexing conditions. Blue curve: DV receiver without the \SI{100}{\giga\hertz} C40 DWDM filter and CV system on. Orange curve: DV receiver with the C40 DWDM filter and CV on. Red curve: DV receiver with the C40 DWDM filter and CV off.}
    \label{fig:influence_CVDV}
\end{figure}

To investigate further the CV--DV leakage, we performed a dedicated background characterization by recording the background count rate with the DV signal blocked under three conditions: (i) DV receiver with the C40 \gls{dwdm} filter and CV transmitter \textit{off}, (ii) with the C40 filter and CV \textit{on}, (iii) DV receiver without the C40 filter and CV \textit{on}. Using the measured background count probabilities in the DV key-rate model, we then simulated the expected DV \gls{skr} as a function of channel attenuation (Fig.~\ref{fig:influence_CVDV}).
As expected, the presence of a narrow \gls{dwdm} filter at the DV receiver is essential to suppress out-of-band light: without the C40 filter the increased background makes the DV \gls{skr} drop to zero already at around \SI{27}{\decibel} of attenuation, while with the filter the DV system remains key-generating up to more than \SI{50}{\decibel}. Furthermore, with the C40 filter in place, the simulated DV \glspl{skr} for the CV on and CV off conditions are essentially indistinguishable, with a 1\% relative difference reached only at \SI{38.2}{\decibel}. At higher losses, a small deviation appears: in the CV on condition, the DV \gls{skr} reaches zero at \SI{51.97}{\decibel}, while in the CV off condition it still gives $\approx\SI{20.8}{\bit\per\second}$ at the same attenuation and only drops to zero at \SI{52.51}{\decibel}. Overall, this confirms that in our configuration (in the range of attenuations we analyzed), the classical CV synchronization sequence and pilot do not introduce any relevant impairment on the DV performance.

\section{Multiplexing over fiber channels~\label{sec:fiber-results}}

\begin{figure}
    \centering
    \includegraphics[width=\columnwidth]{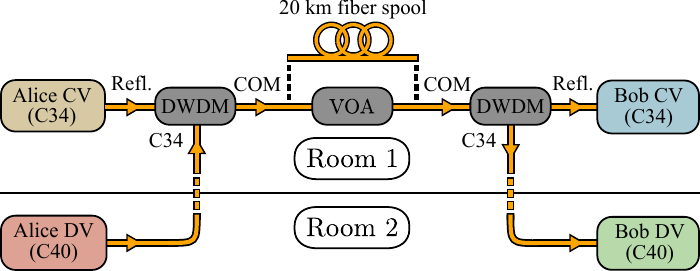}
    \caption{Schematic of the multiplexed \gls{dvqkd} and \gls{cvqkd} setup. The CV transmitter (Alice CV) and receiver (Bob CV) are located in room~1, while the DV transmitter (Alice DV) and receiver (Bob DV) are located in room~2. At the transmitter side, the optical outputs of Alice DV and Alice CV are combined by a 100~GHz DWDM and launched into a common quantum channel, implemented either as a \SI{20}{\kilo\meter} single-mode fiber spool or as a variable-loss link using an electronically controlled VOA. At the receiver side, a second DWDM with identical specifications demultiplexes the two wavelengths and routes them to Bob CV and Bob DV, respectively.}
    \label{fig:multiplexing_setup}
\end{figure}

\begin{figure*}
    \centering
    \includegraphics[scale=0.6]{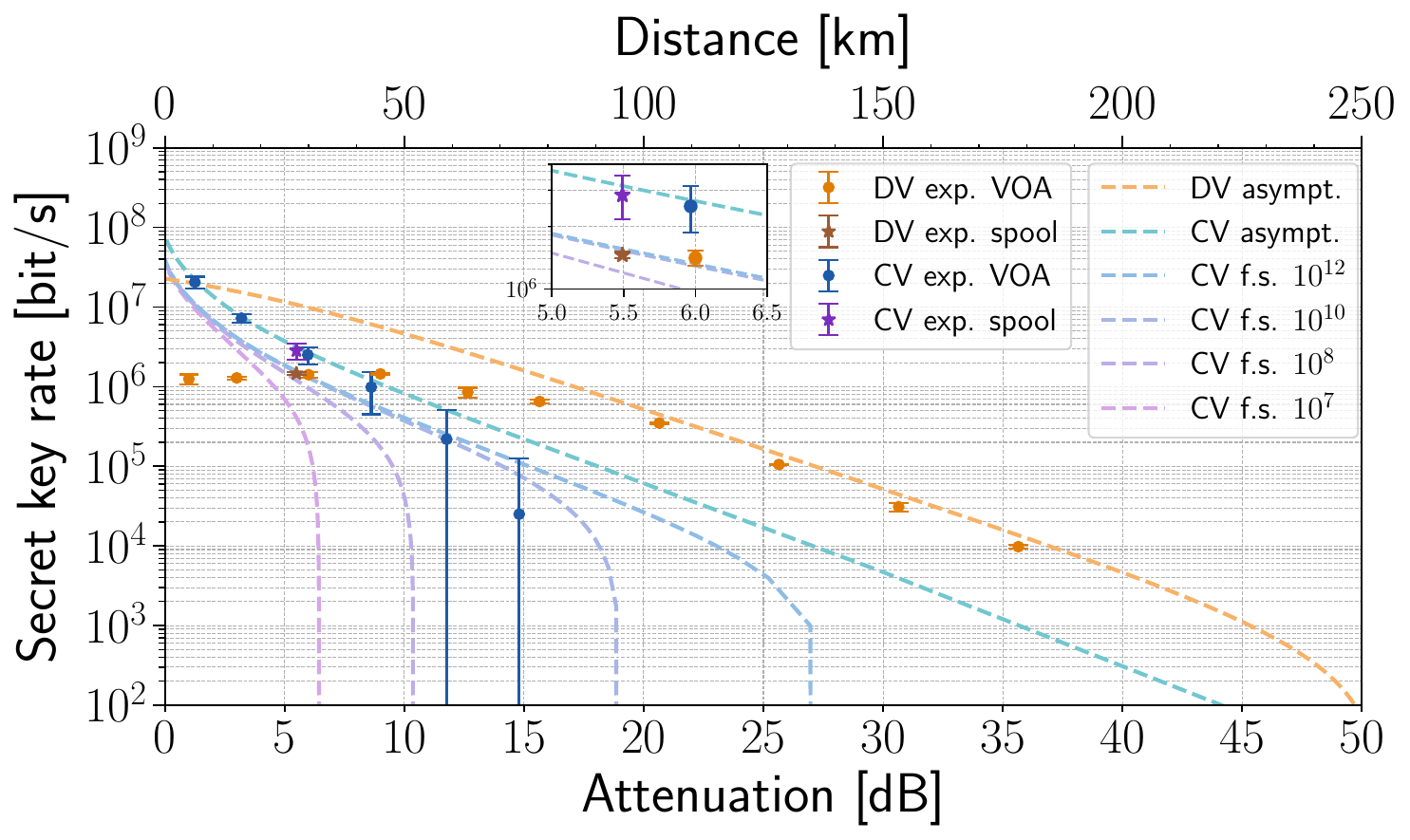}
    \caption{\gls{skr} as a function of channel attenuation when varying the loss with the electronically controlled \gls{voa}, with \gls{dvqkd} and \gls{cvqkd} operating simultaneously on the same optical link. Experimental results are reported as points with error bars in orange for DV-QKD and in blue for CV-QKD; DV data are evaluated in the finite-size regime~\cite{mannalath2025sharp}, while CV data are reported in the asymptotic regime~\cite{fossier_improvement_2009}. 
    Theoretical predictions are reported as dashed lines for both the asymptotic regime and finite-size (f.s.) analysis (for CV-QKD, several block sizes are shown using \cite{leverrier_finite-size_2010}). For each attenuation point, the CV-QKD experimental parameters, namely Alice's modulation variance and the pilot-tone amplitude, were optimized. The points at $\simeq\SI{5.5}{\decibel}$, shown in brown (DV-QKD) and purple (CV-QKD), additionally report the \SI{20}{\kilo\meter} fiber-spool measurements (time-averaged over \SI{60}{\minute}). The upper horizontal axis reports the equivalent fiber link distance, assuming a fiber attenuation coefficient of \SI{0.2}{\decibel\per\kilo\meter}.}
    \label{fig:skr_dv-qkd_voa}
\end{figure*}

A schematic of the multiplexed configuration is shown in Fig.~\ref{fig:multiplexing_setup}. 
The setup combines two independent \gls{qkd} systems operating at distinct wavelengths: a polarization-encoded \gls{dvqkd} transmitter at \SI{1545.32}{\nano\meter} (channel~40) and a coherent \gls{cvqkd} transmitter at \SI{1550}{\nano\meter} (channel~34). 
Each \gls{qkd} system is hosted in a separate laboratory room: the CV transmitter and receiver are located in room~1, while the DV transmitter and receiver are installed in room~2. 
The optical outputs of the two transmitters are combined using a \SI{100}{\giga\hertz}-\gls{dwdm} centered at channel~34, which couples both wavelengths onto the same single-mode fiber channel.
After transmission through either a \SI{20}{\kilo\meter} \gls{smf} spool or an electronically controlled \gls{voa}. After the quantum channel, a second \gls{dwdm} with identical specifications is used to demultiplex the two wavelengths and route them to their respective receivers. 
The configuration with the electronically controlled VOA is used to investigate the simultaneous behavior of both QKD systems as a function of channel attenuation.

\subsection{Variable transmittance}

To investigate the simultaneous operation of the wavelength-division multiplexed \gls{dvqkd} and \gls{cvqkd} systems under varying channel loss, we performed several experiments while changing the attenuation of an electronically controlled \gls{voa} while both systems were running and generating keys at the same time. The resulting \gls{skr} as a function of the channel attenuation is shown in Fig.~\ref{fig:skr_dv-qkd_voa}, where data points with error bars correspond to experimental results and dashed lines to the corresponding theoretical predictions. 
In the plot, the \gls{dvqkd} and \gls{cvqkd} experimental data are shown in orange and blue, respectively, together with the corresponding asymptotic theoretical predictions shown in light orange and light blue.
The \gls{dvqkd} experimental secret key rates are evaluated using a finite-size security analysis~\cite{mannalath2025sharp}, whereas the \gls{cvqkd} experimental rates correspond to the asymptotic regime~\cite{fossier_improvement_2009}. Additional theoretical predictions for \gls{cvqkd} in the finite-size regime~\cite{leverrier_finite-size_2010} are also shown as dashed curves for different block sizes, with colors ranging from light blue to purple.

For low channel attenuation, the \gls{cvqkd} system achieves a higher \gls{skr} than \gls{dvqkd}. For instance, around \SI{1}{\decibel} of loss the CV implementation delivers on the order of $2\times 10^{7}$~bit/s, whereas the DV implementation is limited to about \SI{1.3e6}{\bit\per\second}. As the attenuation increases, the \gls{cvqkd} rate decays more rapidly, while the \gls{dvqkd} system remains operational up to much higher loss and becomes advantageous. 
The crossover point estimated from the experimental data is found at approximately $\SI{7.56}{\dB}$ (roughly \SI{38}{\kilo\meter} of fiber), where both systems achieve a \gls{skr} of $\sim \SI{1.43}{\mega\bit\per\second}$. Beyond this point, the \gls{dvqkd} implementation dominates: for instance, at about $\SI{15.7}{\dB}$ of loss the DV system achieves a \gls{skr} of \SI{6.5e5}{\bit\per\second}, while at \SI{14.8}{\dB} the CV rate has already dropped to $\sim \SI{2.5e4}{\bit\per\second}$.

The discrepancy between the DV experimental data and its theoretical curve at low attenuation is mainly due to limitations on the software used to acquire data from the time-tagger, developed in-house. 
Conversely, the deviation of the CV experimental data from the theoretical prediction, together with the larger fluctuations observed at high attenuation, is attributed to the breakdown of the \gls{dsp} when the \gls{snr} becomes too low, in particular, failures in frame-timing recovery or in clock, frequency, and phase recovery based on the pilot tones. These measurements confirm the expected complementarity: \gls{cvqkd} is more efficient in the low-loss regime, while \gls{dvqkd} extends the operational range to higher attenuation, and both systems can coexist on the same wavelength-multiplexed channel.

\subsection{Fiber spool}

\begin{figure}
    \centering
    \includegraphics[width=\columnwidth]{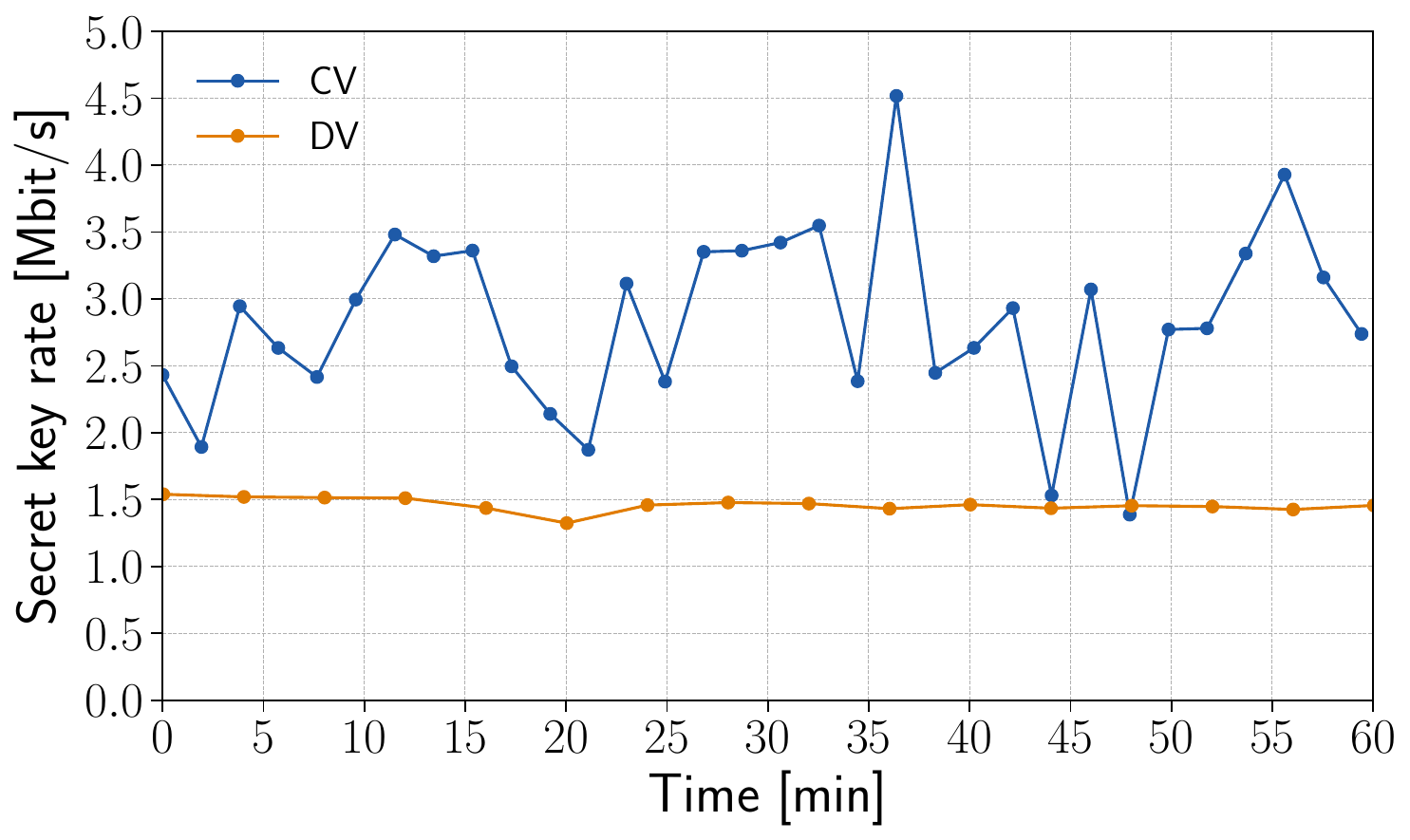}
    \caption{\gls{skr} as a function of time during simultaneous WDM operation of the CV-QKD and DV-QKD systems over a \SI{20}{\kilo\meter} single-mode fiber spool (attenuation $\simeq \SI{5.5}{\decibel}$). The CV-QKD parameters (Alice modulation variance and pilot-tone amplitude) were optimized for the fiber-spool link.}
    \label{fig:SKR_fiber_spool}
\end{figure}

We further validated simultaneous CV--DV operation on a  fiber channel by replacing the variable attenuator with a \SI{20}{\kilo\meter} single-mode fiber spool, corresponding to a channel attenuation of approximately \SI{5.5}{\decibel}. Both \gls{qkd} systems were run simultaneously for about \SI{1}{\hour}, and the resulting \glspl{skr} as a function of time are reported in Fig.~\ref{fig:SKR_fiber_spool}. Overall, both systems sustain Mbit/s key generation over the entire acquisition. The slight monotonic decrease on the \gls{skr} observed for the DV system was attributed to polarization drift within the single mode fiber, as the DV receiver does not count with active polarization compensation. For comparison with the \gls{voa} sweep, the corresponding time-averaged \gls{skr} values using the fiber spool are also reported as a single operating point at $\simeq\SI{5.5}{\decibel}$ in Fig.~\ref{fig:skr_dv-qkd_voa}.



\section{Multiplexing over a free space channel~\label{sec:free-space-results}}

In this phase of the experiment, we investigated the robustness of the quantum systems by integrating their functionalities into a deployed network representative of realistic use cases, such as short-range interconnections between nearby buildings in urban environments (e.g. banks, universities, hospitals). Motivated by the increasing demand for free-space optical links, particularly in scenarios where fiber infrastructure is unavailable or impractical, we demonstrate the feasibility of wavelength multiplexing CV-DV QKD systems within an intermodal network~\cite{Hubel:23}, in which free-space and fiber-based segments are combined to support states propagation through the quantum channel.

\subsection{Testbed description}

\begin{figure*}
    \centering
    \includegraphics[width=0.99\textwidth]{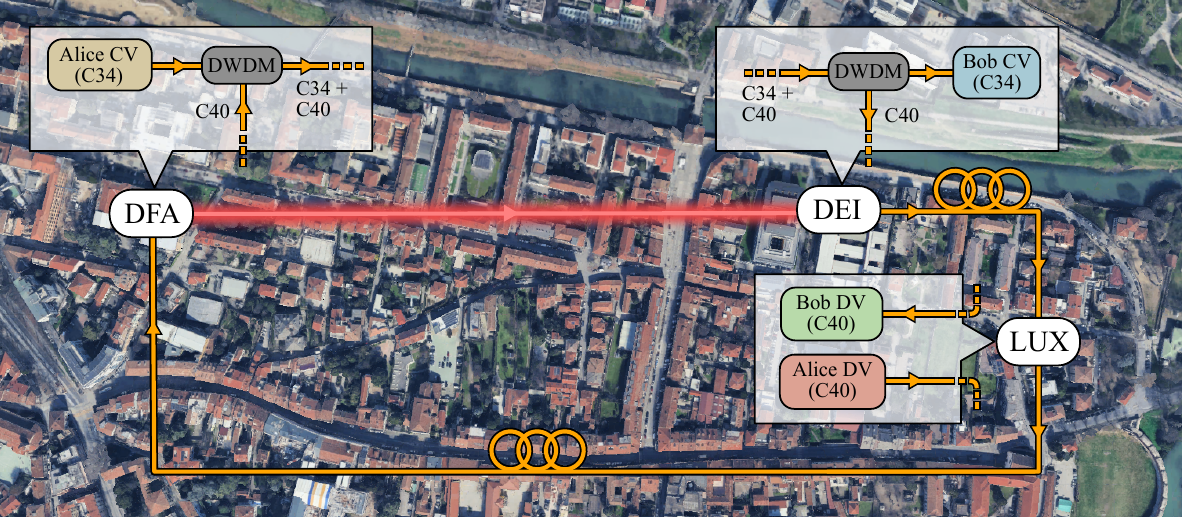}
    \caption{Illustration of the wavelength multiplexed ring topology network in Padova. Data from Google Earth (©2026 Google). The CV system, operating at the wavelength of C34, consists of Alice-CV at DFA, and Bob-CV at DEI. DFA and DEI are connected trhough a free-space channel of \SI{620}{\meter}. The DV system, operating at a wavelenght of C40 is located at LUX. DFA and DEI are connected to LUX through dedicated single mode fibers roughly of \SI{2}{\kilo \meter} and \SI{0.5}{\kilo \meter}.}
    \label{fig:network_architecture}
\end{figure*}

In this testbed, we interconnected multiple buildings of the University of Padova in a ring topology network, with \gls{qkd} units physically deployed at each node to reflect a heterogeneous deployment, as illustrated in Fig.~\ref{fig:network_architecture}.
Alice-CV is installed in a room on the rooftop of the Department of Physics and Astronomy (DFA), while Bob-CV is located in a management room at the Department of Information Engineering (DEI). 
In the case of the DV-system, both the units are hosted in the same laboratory at the Luxor facility (LUX) and are connected to DFA and DEI through a dedicated \gls{smf} links.  
To close the loop, DFA and DEI are interconnected via a free-space optical link. 
Although both systems are intrinsically fiber-based, the inclusion of a free-space segment mimic realistic urban deployment scenarios in which free-space links can be used to bypass fiber out-of-service, and enable rapid network configuration providing redundancy. 
This intermodal configuration allows us to explore the coexistence and performance of CV and DV QKD systems under atmospheric turbulence, offering insight into the robustness and resilience these systems operating in dynamic urban environments.

\begin{figure*}
    \centering
    \includegraphics[width=0.99\textwidth]{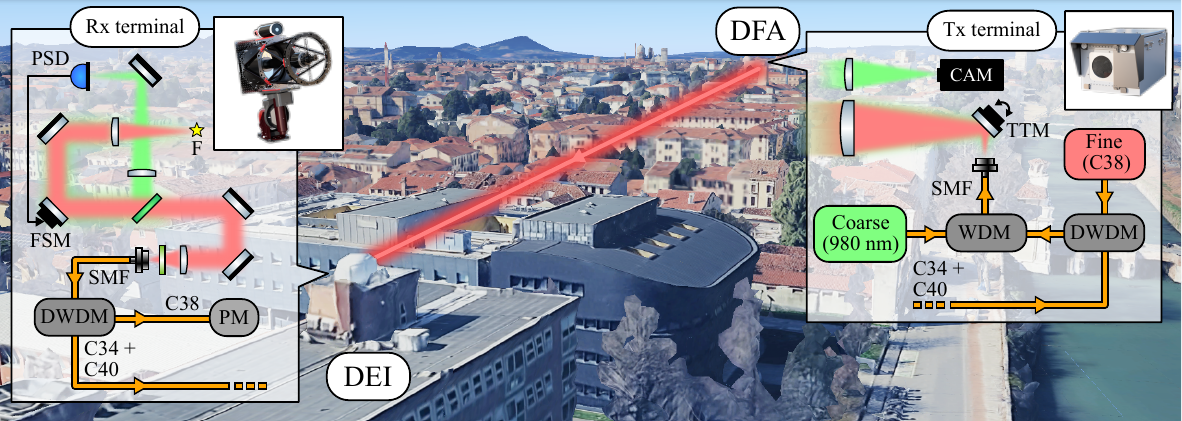}
    \caption{Schematic of the wavelength-division-multiplexed free-space QKD setup. Data from Google Earth (©2026 Google). The Tx terminal, located at DFA, is connected to the Rx terminal, hosted at the DEI optical ground station through a \SI{620}{\meter} free-space optical link. DWDM: dense wavelength-division multiplexer; WDM: wavelength-division multiplexer; SMF: single-mode fiber; TTM: tip-tilt mirror; CAM: camera; F: focal plane; FSM: fast-steering mirror; PSD: position-sensitive device; PM: power meter. }
    \label{fig:free-space-setup}
\end{figure*}

The free-space part of the experiment is carried out over an horizontal urban link of \SI{620}{\meter} connecting the transmitter terminal (Tx), installed on the rooftop of the DFA, with the receiver terminal (Rx), located at the optical ground station at DEI. 
The free-space quantum channel, represented in Fig.~\ref{fig:free-space-setup}, is defined as the propagation from the output of Tx's optical window to the output of the SMF connected to the Rx telescope. 

The experiments begins with Alice-DV unit at LUX generating polarization-encoded optical pulses, which are transmitted to DFA through the dedicated \gls{smf}.
At DFA, the Alice-CV system generates Gaussian-modulated coherent-state optical signals, which are subsequently combined with the incoming quantum signal from the DV system by means of a 100~GHz C34 DWDM filter. 
The multiplexed quantum signals, then propagate along the same \gls{smf} to a second multiplexing stage located internally to the Tx terminal, where additional wavelengths for the free-space segment are added.  
Specifically, the stage is used to add a reference for coarse alignment, implemented using a 980 nm laser, and a monitor of the free-space channel efficiency, provided by measuring the optical power of a \SI{1546.92}{\nano\meter} (C38 of the ITU grid) coupled into the \gls{smf} once at the receiver side.
The resulting multiplexed signals are combined with a \gls{wdm}, and then outputted from the multiplexing stage through a terminated UPC/FC \gls{smf}, connected to a linear stage positioned on the focal plane of the collimation lens, internal to the Tx. The terminal integrates a tip-tilt mirror, allowing for a fine alignment pointing with precision of approximately \SI{2}{\micro \meter}. At the collimation lens, the C-band beams are collimated to a waist of $W_0 = \SI{25}{\milli\meter}$, and sent to free-space through the Tx optical aperture of diameter \SI{150}{\milli \meter}.
The Tx terminal is entirely remotely operated and enables alignment with the Rx by means of an alt-azimuth mount and a camera (CAM) with a full-angle field of view (FOV) of $8 \times \SI{6.4}{\milli\radian\squared}$.

At the receiver side, an $f/8$ optical telescope is used as Rx. 
The terminal collects the transmitted beam, and using the information of the beacon laser detected on a position-sensitive-device (PSD), it feedback a fast-steering mirror (FSM) to couple the C-band signals into a \gls{smf}, as done in Ref. \cite{bolanos_ghz-rate_2026}. 
Once these signals are injected, a first C38 \gls{dwdm} separate the monitor wavelength from the quantum signals, and send it to a power meter for the continuous estimate of the free-space channel transmittance. 
For the quantum signals, a subsequent C34 \gls{dwdm} separates the CV channel from the DV, which is forwarded back to LUX, where it reach Bob~DV for the measurement.

The access fiber from LUX to the Tx terminal (DFA) contributes \SI{2.75}{\dB}, while the return fiber from the Rx terminal (DEI) to LUX contributes \SI{2.59}{\dB}. At the Tx terminal, a multiplexing stage combines the \gls{cvqkd} channel (C34) and the \gls{dvqkd} channel (C40) with the free-space monitoring channel (C38) and a 980-nm beacon. Overall, the corresponding \gls{wdm} filters contributes \SI{3.64}{\dB} of insertion loss, and the telescope-fiber interface adds \SI{0.7}{\dB} of geometrical loss. At the receiver, the demultiplexing \gls{wdm} stage (separating C38/C34/C40) introduces an additional \SI{0.93}{\dB} insertion loss. These terminal losses are treated as trusted in the CV-QKD model but untrusted in DV-QKD. Therefore, DV experiences an additional overall attenuation
\begin{equation}
\Delta \eta_{\mathrm{DV-CV}} = \SI{10.61}{\dB}
\end{equation}
due to the additional fiber links and untrusted terminal insertion losses. Although the systems are deployed in an intermodal network, our analysis focuses on the free-space channel.

Once again, we emphasize that the \gls{cvqkd} implements the local local oscillator technique which prevents attacks with respect the transmitted local oscillator technique. While this technique has matured for fiber-based channel, it has, up to now, only been deployed on emulated free-space channel, and this work hence also represents an advance with respect to the state of the art in this regard~\cite{liao_high-rate_2025}. The system also provides a secret key rate in the $\sim$ Mbit/s range, where previous demonstrations, even using the transmitted local oscillator technique achieved in the order of hundreds of kbit/s~\cite{zheng_free-space_2025}.

\subsection{Free-space results}

\begin{figure*}[t]
    \centering
    \subfloat[\label{fig:skr_dv-qkd_free-space}]{
        \includegraphics[width=0.485\textwidth]{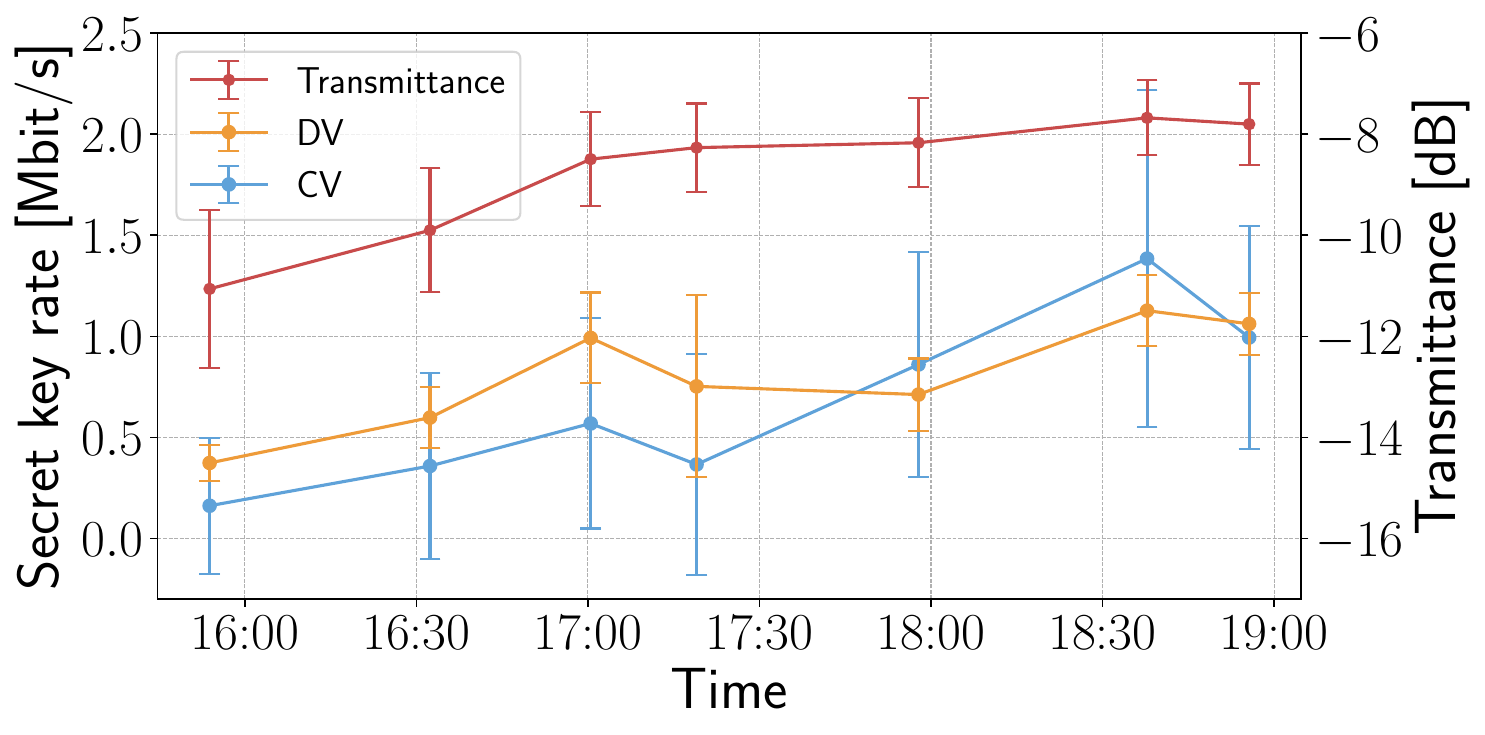}
    }\hfill
    \subfloat[\label{fig:meteo}]{
        \includegraphics[width=0.485\textwidth]{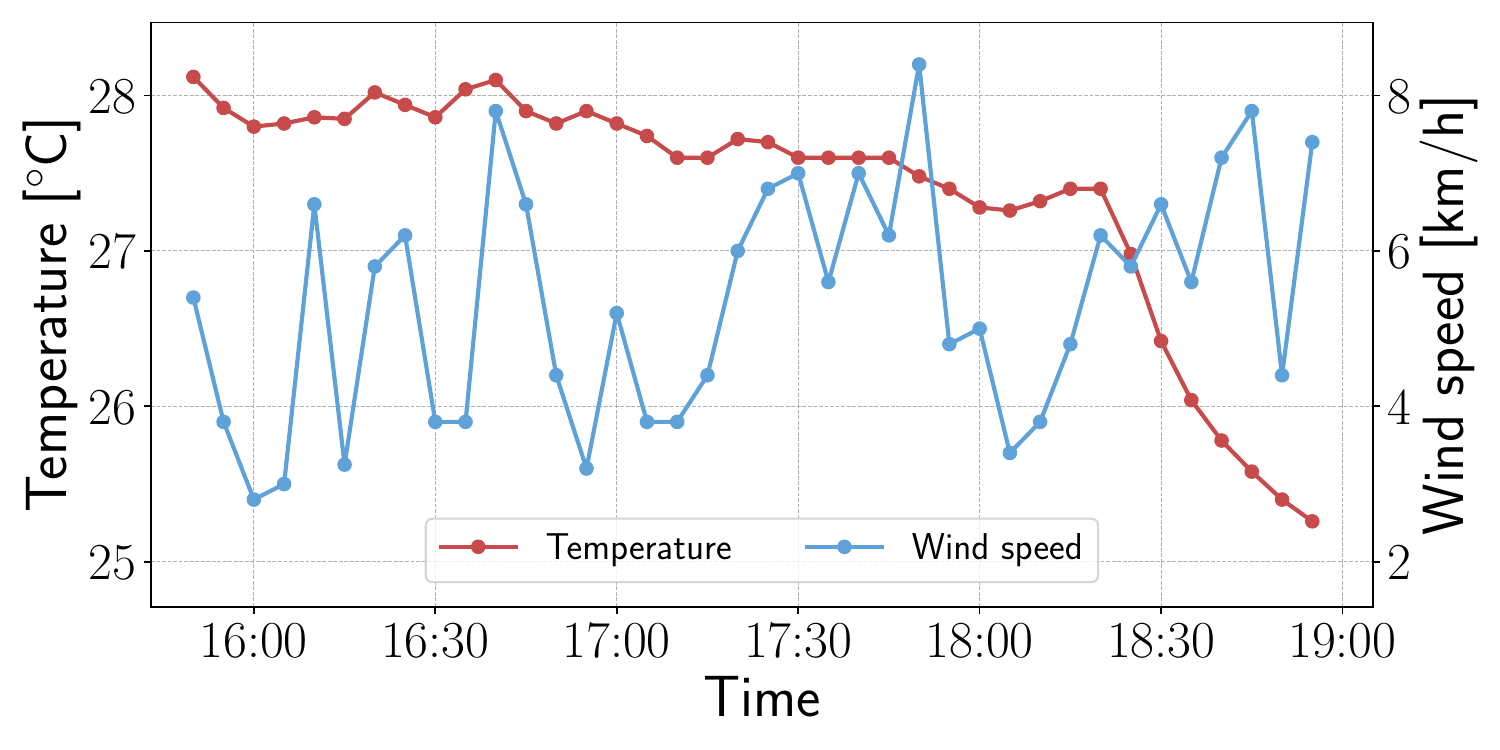}
    }
    \caption{Daylight free-space run on September, 15th 2025 (Padova, local time, UTC+2; sunset at $\sim$19{:}26).
    \textbf{(a)} \gls{skr} as a function of time during simultaneous operation of wavelength-division-multiplexed CV-QKD and DV-QKD. The DV and CV secret key rates are shown in orange and light blue, respectively, while the red curve shows the measured free-space channel transmittance. The DV-QKD path includes an additional fixed loss of $\sim\SI{10.6}{\decibel}$ from the fiber segments linking LUX to the Tx/Rx terminals and from terminal insertion losses (not included in the red curve).
    \textbf{(b)} Meteorological conditions over the same time interval, recorded by the DEI weather station at the Department of Information Engineering (University of Padova). Measurements are reported as 5-min averages of temperature (red) and wind speed (light blue).}
    \label{fig:free-space_skr_and_meteo}
\end{figure*}

The experiment has been performed in daylight during the afternoon of September 15th, 2025 for a duration of 3 hours. The secret key rates of the wavelength-division multiplexed free-space link as a function of time are shown in Fig.~\ref{fig:skr_dv-qkd_free-space}. During the experiment, the estimated free-space attenuation (red curve, right axis) progressively decreases from approximately $(11.1 \pm 1.6)\,\mathrm{dB}$ at the beginning of the measurement to $(7.8 \pm 0.8)\,\mathrm{dB}$ toward the end. This trend is accompanied by a stabilization of the free-space link starting at around 18:30, as evidenced by the reduced fluctuations in the attenuation. The transition coincides with a rapid temperature drop, occurring when the sun begins to move behind the surrounding hills. Consistently, the attenuation uncertainty, represented by the error bars, is larger at the beginning of the experiment and progressively decreases over the course of the measurement.

The orange and light-blue curves report the experimentally measured secret key rates of the multiplexed \gls{dvqkd} and \gls{cvqkd} systems, respectively, with error bars corresponding to one standard deviation over the samples collected in each time interval. In the first part of the acquisition, when the free-space losses are higher, the \gls{dvqkd} system achieves a larger \gls{skr} than the \gls{cvqkd} one. As the channel attenuation decreases, the \gls{cvqkd} performance improves and eventually overtakes the \gls{dvqkd} rate. Over the run, we compute a weighted time average of the secret key rates. This yields averaged secret key rates of
$R_{\mathrm{DV}} = (0.838 \pm 0.003)\,\mathrm{Mbit/s}$ for the \gls{dvqkd} system and
$R_{\mathrm{CV}} = (0.70 \pm 0.07)\,\mathrm{Mbit/s}$ for the \gls{cvqkd} system, where the uncertainties correspond to the standard error of the weighted mean. During the run we obtain an average channel loss of
$L_{\mathrm{FS}} = (8.73 \pm 0.01)~\mathrm{dB}$.
We also report the peak values observed in the raw \gls{skr} samples during the run: \SI{2.785}{\mega\bit\per\second} for \gls{cvqkd} (around 18{:}42) and \SI{1.920}{\mega\bit\per\second} for \gls{dvqkd} (around 17{:}21).

These results show that both systems can sustain Mbit/s secret key rates under realistic daylight conditions on an urban free-space link, despite the combined effects of background light and turbulence.

\section{Conclusion~\label{sec:conclusion}}

We have experimentally demonstrated wavelength-division multiplexing of independent \gls{cvqkd} and \gls{dvqkd} systems operating simultaneously over the same optical link, in both fiber and free-space. Importantly, we observe no measurable multiplexing-induced penalty within the explored losses regime, simultaneous operation does not produce any detectable increase in DV-QKD QBER or CV-QKD excess noise. Consistently, a dedicated background characterization shows that, up to $\sim 38$~dB of channel attenuation, the simulated DV-QKD \gls{skr} is essentially unchanged during simultaneous CV–DV operation.

On fiber, measurements using an electronically controlled \gls{voa} as a variable-loss channel confirm the expected complementarity between the two protocols: \gls{cvqkd} achieves higher \gls{skr} at low loss, while \gls{dvqkd} becomes advantageous at higher attenuation. The experimental crossing occurs around $7.56~\mathrm{dB}$, where both systems reach $\simeq 1.43~\mathrm{Mbit/s}$. We further validated simultaneous operation on a \SI{20}{\kilo\meter} fiber spool, where both systems sustained Mbit/s secret key rates for one hour of acquisition. Finally, we deployed the WDM coexistence in an urban intermodal network, combining fiber links with a \SI{620}{\meter} daylight free-space link, and demonstrated Mbit/s key generation under time-varying atmospheric attenuation, including free-space \gls{cvqkd} operation with local local oscillator.

These results show that hybrid CV--DV architectures can be implemented on shared fiber and free-space infrastructure with negligible impact from coexistence, and support their adoption in future quantum communication networks where high-throughput, short-range users and long-reach, high-loss links must be served simultaneously on the same physical channel.

\acknowledgments

This work was supported by European Union’s Horizon Europe research and innovation program under the project Quantum Secure Networks Partnership (QSNP), grant agreement No 101114043. Views and opinions expressed are however those of the authors only and do not necessarily reflect those of the European Union or European Commission-EU. Neither the European Union nor the granting authority can be held responsible for them.

\section*{Author contributions}
ER, FV and PV designed and implemented the free space link. TL, YP and ED designed, and implemented the CV-QKD system. MRB, MA, GV and PV designed and implemented the DV-QKD system. Data acquisition and system operation was performed by MS, MB, ER, and YP. Data analysis was performed by MS, MRB and YP. Supervision was provided by GV, PV, YP and MA. All authors participated in the discussion of results and contributed to the current manuscript.

\bibliography{bibliography}
\newpage
\onecolumngrid
\appendix

\section{CV-QKD Digital Signal Processing}\label{app:dsp}

The CV-QKD transmitter prepares Gaussian-modulated coherent-state signals using the IQ modulator driven by DAC-synthesized electrical waveforms. The Gaussian distributed symbols are pulse-shaped with a root-raised-cosine (RRC) filter (roll-off $\beta_{\mathrm{RRC}}=0.3$) at a symbol rate $R_s=\SI{100}{\mega\baud}$. A digital frequency shift $f_{\text{shift}}=\SI{170}{\mega\hertz}$ is applied to translate the quantum band away from DC and mitigate low-frequency technical noise. Each frame starts with a maximum-length sequence generated using a linear-feedback shift register with \(n_{\mathrm{LFSR}}=16\) bits, thus producing a pseudo-random binary sequence of length \(L_{\mathrm{MLS}}=2^{n_{\mathrm{LFSR}}}-1\) with near-ideal autocorrelation properties. The sequence is used for accurate frame detection and timing alignment in noisy environments.
Two pilot tones at $f_{\text{pilot},1}=\SI{400}{\mega\hertz}$ and $f_{\text{pilot},2}=\SI{420}{\mega\hertz}$ are frequency-multiplexed with the quantum signal and used for clock recovery as well as carrier-frequency and phase tracking. The waveform is produced by the DAC at \SI{2}{\giga\sample\per\second} and applied to the IQ modulator.

No external reference clock is distributed between Alice and Bob, the resulting sampling-clock mismatch is recovered from the pilot tones in \gls{dsp}, while frame timing is obtained from the maximum-length sequence preamble.
Alice and Bob use independent CW lasers with a controlled small relative frequency offset (sub-GHz scale offset), compensated in \gls{dsp} using the pilot tones.

On Bob's side, the signal from the balanced detector is acquired using an ADC at \SI{2.5}{\giga\sample\per\second}, with an acquisition window of $T_{\mathrm{acq}}=\SI{70}{\milli\second}$ per frame. The \gls{dsp} performs synchronization, matched filtering, pilot-aided frequency/phase recovery, and symbol extraction prior to parameter estimation. More information on the \gls{dsp} can be found in~\cite{Pietri2024}.

The main \gls{dsp} and sampling parameters are summarized in Table~\ref{tab:dsp_params}.

\begin{table}[h]
    \centering
    \begin{tabular}{c|c} 
       \textbf{Parameter} & \textbf{Value} \\
        \hline
        $\beta_{\mathrm{RRC}}$    & 0.3 \\
        $R_s$                     & 100 MBaud \\
        $f_{\text{shift}}$        & 170 MHz \\
        $f_{\text{pilot},1}$      & 400 MHz \\
        $f_{\text{pilot},2}$      & 420 MHz \\
        $n_{\mathrm{LFSR}}$  & 16 \\
        $T_{\mathrm{acq}}$          & 70 ms \\
        DAC rate                  & 2 GSa/s \\
        ADC rate                  & 2.5 GSa/s \\
        Modulation                & Gaussian \\
    \end{tabular}
    \captionsetup{labelformat=empty}
    \caption{Values of the main \gls{dsp} and sampling parameters used in the CV-QKD setup. Here $\beta_{\mathrm{RRC}}$ is the roll-off factor of the root-raised-cosine pulse-shaping filter, $R_s$ is the symbol rate, $f_{\text{shift}}$ is the digital frequency shift applied to the quantum data, $f_{\text{pilot},1}$ and $f_{\text{pilot},2}$ are the frequencies of the pilot tones used for clock recovery and pilot-aided carrier-frequency and phase tracking, 
    $n_{\mathrm{LFSR}}$ is the number of bits in the linear-feedback shift register used to generate the maximum-length synchronization sequence, $T_{\mathrm{acq}}$ is the acquisition window per frame, and the DAC/ADC rates are the sampling rates of the generation and detection hardware.}
    \label{tab:dsp_params}
\end{table}

\end{document}